# Target Detection via Quantum Illumination: Range Equation

Hossein Allahverdi[1*], M.H. Qamat[2], M. Nowshadi[3]

[1] *Laser and Plasma Research Institute, Shahid Beheshti University, Tehran, Iran.*

[2] *School of Electrical & Computer Engineering, Shiraz University, Shiraz, Iran.*

[3] *Physics Department, Vali-e-Asr University of Rafsanjan, Kerman, Iran.*

## Abstract

In this article, the basic principle of target detection based on Gaussian state quantum illumination (QI) has introduced. The performance of such system has compared with its classical counterpart, which employs the most classical state of light, i.e., coherent state, to illuminate the target region. By deriving the maximum range equation, we have demonstrated that the quantum illumination based target detection system is especially advantageous at low transmission powers, which make these systems suitable for short range applications like biomedical imaging or covert detection schemes.

## I.  INTRODUCTION

Quantum technology is an emerging field that employs quantum phenomena, like quantum tunneling or quantum entanglement, to boost the performance of the systems beyond the best classical strategies. Until now, the advantages of quantum technologies in information processing [1-3], computation [4,5], communication systems [6-8], and measurement [9] have been demonstrated. Therefore, standoff sensing is not an exception and there are several approaches to employ quantum mechanical effects in order to enhance the performance of such systems [10]. Quantum illumination (QI) is a quantum optical sensing technique, which uses quantum entanglement to improve the detection of a low-reflective object that is present in a lossy and noisy environment [11-13].

In QI strategy, which was first proposed by Lloyd [11] , a pair of entangled signal and idler photons used to illuminate the target, such that the signal transmits to probe the environment, while the idler retained in the transmitter until the signal photon returns; Finally, a joint measurement on the returning signal together with the retained idler take place to distinguish the signal photons from the thermal noises that comes from the [11,12]. The first experimental realization of QI in optical domain have been done by Lopaeva *et.al.* [14]. The need for joint measurement in this approach limits the use of QI in ranging applications. However, one can overcome this limitation by employing digital receivers to do separate measurements on the returned signal and retained idler, and then correlate the two measurement results using match filtering in order to distinguish the signal from the background noise [15-16].

---

[*] h.allahverdi@mail.sbu.ac.ir

Despite the numerous efforts to use the idea of QI in order to enhance the sensitivity of target detection in microwave domain [17-21], there is no comprehensive study has occurred in order to determine the maximum achievable range of this scheme and the advantages over its classical counterpart. In this paper, we are going to derive an explicit equation for the maximum range of the QI based target detection system and compared the results with the radar range equation of the classical radars. In this study, we assumed a pair of entangled signal and idler photons obtained from the continuous wave spontaneous parametric down conversion (CW-SPDC) process [12]. Moreover, we have supposed that the classical illumination (CI) system utilizes two correlated coherent states of electromagnetic (EM) field as signal and idler photons, which can obtained by dividing a single coherent EM field through the power divider (for microwave EM field) or beam splitter (for optical EM field) [14]. In each case, the covariance matrix of the signal and idler beams have calculated, which quantifies the amount of correlations between the signal and idler beams. Comparison between the results of the QI strategy to the classical one reveals that the QI based target detection is especially advantageous at low transmission powers, which make these systems suitable for short range applications like biomedical imaging or covert detection schemes.

The paper is organized as follows. Sec.II is devoted to description of the quantum and classical correlated signal and idler beams. In Sec.III, the range equation for the QI based target detection system have derived and compared with the radar range equation of the classical radar system. Finally, the results and conclusions are in SecIV.

## II.  QUANTUM VS CLASSICAL ILLUMINATION

### A. Gaussian state quantum illumination

Consider a pair of entangled signal and idler photons generates through the CW-SPDC process. Quantum mechanical representation of such entangled pairs in terms of number state of the signal and idler fields is [22]

$$|\psi\rangle_{SI} = \sum_{n=0}^{+\infty} \sqrt{\frac{N_s^n}{(N_s+1)^{n+1}}} |n\rangle_S |n\rangle_I \qquad (1)$$

where $N_s$ is the mean photon number per mode. The covariance matrix that quantifies the amount of correlation between the signal and the idler modes obtained as

$$E_{SI}^{(QI)} = \begin{pmatrix} R_{SS}^{(QI)} & R_{SI}^{(QI)} \\ R_{IS}^{(QI)} & R_{II}^{(QI)} \end{pmatrix} \qquad (2)$$

in which $R_{jk}^{(QI)}$s are $2 \times 2$ blocks that given by

$$R_{jk}^{(QI)} = \begin{matrix} \langle \hat{I}_j \hat{I}_k \rangle & \langle \hat{I}_j \hat{Q}_k \rangle \\ \langle \hat{Q}_j \hat{I}_k \rangle & \langle \hat{Q}_j \hat{Q}_k \rangle \end{matrix} \qquad (3)$$

Here $\langle \cdots \rangle$ indicates the quantum mechanical expectation value over the entangled Gaussian state given by Eq.(1), and the operators $\hat{I}_S$, $\hat{Q}_S$, $\hat{I}_I$ and $\hat{Q}_I$ are Hermitian operators represent the in-phase and quadrature voltages of the signal (with index 'S') and idler (with index 'I') beams, respectively. These operators can be written in terms of the annihilation and creation operators of the signal and idler modes, i.e. $\hat{a}_S$ ($\hat{a}_S^\dagger$) and $\hat{a}_I$ ($\hat{a}_I^\dagger$), as $\hat{I}_S = (\hat{a}_S + \hat{a}_S^\dagger)/\sqrt{2}$, $\hat{Q}_S = (\hat{a}_S - \hat{a}_S^\dagger)/i\sqrt{2}$, $\hat{I}_I = (\hat{a}_I + \hat{a}_I^\dagger)/\sqrt{2}$ and $\hat{Q}_I = (\hat{a}_I - \hat{a}_I^\dagger)/i\sqrt{2}$ [22]. Therefore, after straightforward calculations, the elements of the covariance matrix for the entangled Gaussian state can be obtained as

$$R_{ss}^{(QI)} = R_{ii}^{(QI)} = \begin{matrix} S & 0 \\ 0 & S \end{matrix} \qquad (5a)$$

$$R_{si}^{(QI)} = R_{is}^{(QI)} = \begin{matrix} C_q & 0 \\ 0 & -C_q \end{matrix} \qquad (5b)$$

with $S = 2N_s + 1$, and $C_q = 2\sqrt{N_s(N_s + 1)}$. In following, we continue our analysis by calculating the covariance matrix of a correlated coherent states of an EM field.

## B. Coherent state classical illumination

We have considered a pair of correlated coherent state of an EM field, which serve as a signal and idler beams, as a reference to compare the performance the QI strategy with. One can generate such classically correlated states by dividing a coherent state of EM field through a power divider (for microwave EM field) or beam splitter (for optical EM field) [14]. Quantum mechanical representation of these states is given by

$$|\psi\rangle_{CI} = |\alpha\rangle_S \otimes |\alpha\rangle_I \qquad (6)$$

where $|\alpha\rangle_S$ and $|\alpha\rangle_I$ are coherent states of EM field that represents the signal and idler beams, respectively. These states can be written in terms of number states of the signal and idler fields as

$$|\alpha\rangle_k = e^{-|\alpha|^2/2} \sum_{n=0}^{+\infty} \frac{\alpha^n}{\sqrt{n!}} |n\rangle_k \qquad (7)$$

With $k = S, I$ and $|\alpha|^2 = N_s/2$ as the mean photon per mode. Using Eq.(6) and Eq.(7), one can obtain the covariance matrix for the correlated coherent states of EM field as

$$E_{SI}^{(CI)} = \begin{pmatrix} R_{SS}^{(CI)} & R_{SI}^{(CI)} \\ R_{IS}^{(CI)} & R_{II}^{(CI)} \end{pmatrix} \qquad (8)$$

in which

$$R_{SS}^{(CI)} = R_{II}^{(CI)} = \begin{matrix} S & 0 \\ 0 & S \end{matrix} \qquad (9a)$$

$$R_{SI}^{(CI)} = R_{IS}^{(CI)} = \begin{matrix} C_c & 0 \\ 0 & -C_c \end{matrix} \qquad (9b)$$

with $S = 2N_s + 1$ and $C_c = 2N_s$.

Comparing the off-diagonal elements of the covariance matrix for classical and quantum correlated signals leads to

$$\frac{C_c}{C_q} = \left(1 + \frac{1}{N_s}\right)^{-1/2} \tag{10}$$

As $N_s > 0$, so we can conclude that $C_c/C_q$ is always lower than one. In Figure 1, this parameter have plotted versus the mean generated photon pairs per mode. This figure reveals that quantum transmitter allows the generation of correlations that is beyond the best possible classical correlations. Moreover, it shows the quantum advantage is especially significant at low $N_s$, i.e. at low transmission powers, and it gradually increase as the transmission power grows. For example, according to this figure, a quantum transmitter with $N_s = 0.5$ generates correlations that is about 40% greater with respect to a best classical transmitter. Note that the transmission power $P_t$ of a transmitter that generates $N_s$ photon per mode at frequency $f$ and bandwidth $B$ is obtained as

$$P_t = N_s h f B \tag{11}$$

where $h = 6.63 \times 10^{-34}$js is a Planck constant. According to this equation, the output power of a signal transmitter with $N_s = 0.5, = 7\text{GHz}$ and $B = 1\text{GHz}$ equals $-116.34\text{dBm}$.

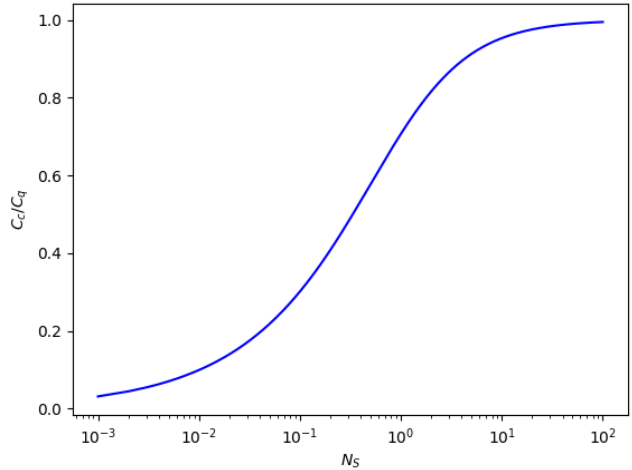

Figure 1. The ratio of the classical correlation coefficient to the quantum correlation coefficient, versus the mean transmitted photon pairs per mode.

### III. RANGE EQUATION

In this section, we are going to derive an explicit expression for the maximum achievable range of the QI-based target detection system introduces in Sec.II.

## A. Maximum range of a classical target detection system

We begin our analysis by derivation of the transmission coefficient of transmitter-to-target-to-receiver channel. This parameter defined as a ratio of the received power to the transmitted power, i.e., $\eta = P_r/P_t$ with $P_t$ and $P_r$ are the transmitted and received power, respectively. According to [23], for a monostatic radar system with a signal generator of power $P_t$, the power of the received signal $P_r$ is given by

$$P_r = \frac{\sigma G A P_t F^2}{(4\pi)^2 R^4} \tag{12}$$

where $\sigma$ is the target RCS, $G$ and $A$ are antenna gain and aperture, respectively, and relates to each other by $G = 4\pi A/\lambda^2$ with $\lambda$ the signal wavelength. Moreover, $R$ is a range to the target and $F$ is a form factor that represents the loss of a signal due to the atmospheric absorption, scattering and etc. In general, $F$ is a function of the signal frequency $\nu$ and the range to the target $R$ and obtained as [24]

$$F = 10^{-\gamma R/10} \tag{13}$$

where $\gamma$ is the absorption coefficient and depends on the signal frequency, temperature and density of the atmosphere [24]. In our analysis, we only consider signal loss due to the atmospheric absorption. The $\gamma$ for different frequencies is selected according to Figure 2, which is obtained from spectroscopic data of $O_2$ and $H_2O$ molecules provided by International Telecommunication Union-Radiocommunication (ITU-R). According to Eq.(12) The transmission coefficient of the channel obtained as

$$\eta = \frac{\sigma G A F^2}{(4\pi)^2 R^4}. \tag{14}$$

As it is well known, the signal to noise ratio (SNR) of a radar system is given by

$$SNR = \frac{P_r}{P_B} \tag{15}$$

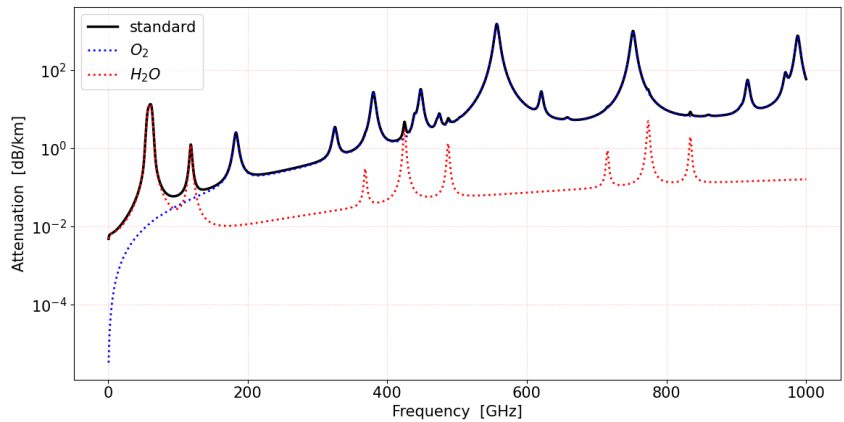

Figure 2. Spectrum of the atmospheric attenuation coefficient ($\gamma$), which is obtained from the spectroscopic data of $O_2$ and $H_2O$ provided by International Telecommunication Union-Radiocommunication (ITU-R).

where $P_B = k_B T_{eff} B = N_B h \nu B$ is the total thermal noise power (due to the antenna noise and receiver noise temperature $T_{eff}$) and $P_r$ is the signal power at the receiver input. Here $N_B$ is the mean thermal photon number per mode and obtained as $N_B = k_B T_{eff}/h\nu$ with $k_B = 1.38 \times 10^{-23} \text{jK}^{-1}$ the Boltzmann constant. Using Eq.(11) and Eq.(15), one can easily obtain the signal to noise ratio as $SNR = \eta N_s/N_B$. Considering lossless coherent integration over $M$ independent and identically distributed measurements, an effective signal to noise ratio obtained as [23]

$$SNR_{eff} = \frac{\eta M N_s}{N_B} \tag{16}$$

As mentioned in [10,23], There is a minimum value of $SNR$ that allows the detection of a signal at the detector output. That is to say target detection is declared only when $SNR_{eff}$ exceeds a certain threshold $SNR_{min}$, i.e., $SNR_{eff} > SNR_{min}$. Minimum SNR is depended to the probability of detection and false alarm probability at the detectors output. For example, for $P_d = 0.7$ and $P_{fa} = 10^{-6}$, the minimum signal to noise ratio is $SNR_{min} = 10dB$. This imposes an upper limit on the detection range for the detection system. Using Eq.(14) and Eq.(16), the maximum range $R_{max}$ of the detection system is obtain as

$$R_{max} = \left( \frac{\sigma G A M N_s F^2}{(4\pi)^4 N_B SNR_{min}} \right)^{1/4} \tag{17}$$

### B. Maximum range of a QI based target detection system

As mentioned in Eq.(10) and illustrated in Figure 1, quantum transmitters that generates Gaussian state entangled photons are able to produce correlations stronger than any classical transmitters with equal transmit power. According to [14] such quantum transmitters improves target detection sensitivity by a factor of $1 + 1/N_s$ rather than a classical counterpart. In other words, the signal to noise ratio for a target detection system equipped with a quantum transmitter is $1 + 1/N_s$ times greater that a classical target detection system with the same power. That is to say the threshold signal to noise ratio for a QI based target detection system is $1 + 1/N_s$ times lower than that for a classical system, i.e., $SNR_{min}^{(QI)} = (1 + 1/N_s)^{-1} SNR_{min}$. Substituting this result in Eq.(17) leads to the maximum range equation of a QI based target detection system, which is

$$R_{max}^{(QI)} = \left(1 + \frac{1}{N_s}\right)^{1/4} \left( \frac{\sigma G A M N_s F^2}{(4\pi)^4 N_B SNR_{min}} \right)^{1/4} \tag{18}$$

Maximum range of a QI- and CI-based target detection system has compared In Figure 3 for signal frequencies 7, 95 and 1000 GHz. Simulation parameters are in Table 1. Note that in this simulation we have assumed that the effective antenna area is equal to 0.5 and G is calculated for different frequencies. In this figure, the maximum range of these systems plotted versus the generated photons per mode $N_s$. As illustrated in this figure, the maximum range for both QI- and CI-based systems increases as the transmission power rises, but the advantage of the Quantum system is especially significant at low powers. Moreover, This figure shows that QI-based target detection system with signal frequency $f = 1THz$ allows detection of targets at range 400m, when it transmits only $10^{-2}$ photons per mode, which is equivalent to transmit power of $-111.79dBm$. While the classical system with the same transmit power is able to detect targets at range 120m.

Table 1. System parameters

| Parameter | Notation | Value |
|---|---|---|
| Effective antenna area [m²] | $A$ | 0.5 |
| Noise power [dBm] | $P_B$ | -63.82 |
| Integration time [s] | $\tau$ | 1 |
| Bandwidth [GHz] | $B$ | 1 |
| Minimum signal to noise ratio at receiver [dB] | $SNR_{min}$ | 10 |
| Probability of detection | $P_d$ | 0.7 |
| False alarm probability | $P_{fa}$ | $10^{-6}$ |
| Target cross section [m²] | $\sigma$ | 1 |
| Signal frequencies [GHz] | $f$ | 7-95-1000 |

## IV. CONCLUSION

In summary, we have investigated the performance of a QI-based target detection system and derived and explicit equation for the maximum achievable range of this system to detect the target. In addition, we compare the results with the best possible classical strategy, in which a pair of classically correlated coherent state EM fields employed as a signal and idler beams to detect a target. We showed that using quantum transmitters, which generates a pair of Gaussian state entangled photon pairs, in target detection system can improve the detection sensitivity and the maximum range of detection rather than a best classical strategy with the same transmit power. We have also demonstrated that the quantum strategy is especially advantageous at low transmission powers. This is due to the fact that the greatness of Gaussian state entangled beams correlation compare to that of coherent state correlated beams is significant at low transmit powers; i.e., As the transmit power of these quantum transmitters rises, correlation of the entangled beams reduces to the correlation of the classically correlated ones. This feature restricts the applications of a QI-based target detection systems to short range applications like biomedical imaging or short range covert detection systems. It seems that quantum illumination is not a proper strategy for enhance sensitivity of target detection for long range applications.

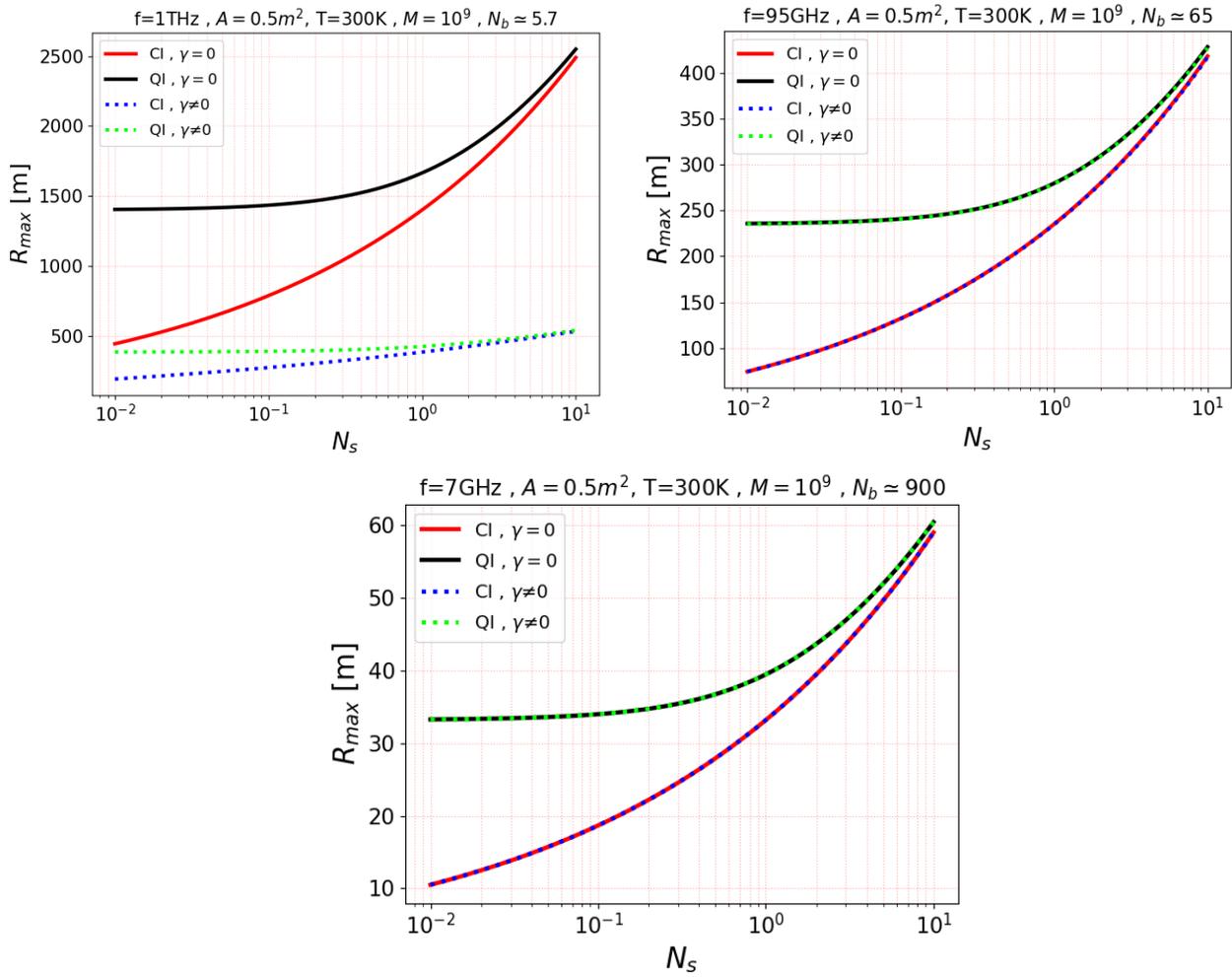

Figure 3. Maximum range of a QI- and CI-based target detection system versus the mean generated photon per mode at frequencies of $7GHZ$, $95GHz$ and $1THz$. The other system parameters are given in table 1.